\documentclass[journal]{IEEEtran}
 
\usepackage{amsmath} 
 
\usepackage{cite}

\usepackage{tikz} 
\usetikzlibrary{decorations.pathreplacing,decorations.pathmorphing,shapes,backgrounds,patterns,decorations.text}

\newcommand{\arrayantenna}[2]{
\draw[very thick,-] (#1,#2) -- (#1+1.0,#2);
\draw[very thick,-] (#1,#2) -- (#1,#2+.2);
\draw[very thick,-] (#1+.2,#2) -- (#1+.2,#2+.2);
\draw[very thick,-] (#1+.4,#2) -- (#1+.4,#2+.2);
\draw[very thick,-] (#1+.6,#2) -- (#1+.6,#2+.2);
\draw[very thick,-] (#1+.8,#2) -- (#1+.8,#2+.2);
\draw[very thick,-] (#1+1.0,#2) -- (#1+1,#2+.2);

\draw[very thick,-] (#1.25,#2-.5) -- (#1+.5,#2);
\draw[very thick,-] (#1.75,#2-.5) -- (#1+.5,#2);
 
}
 
\newcommand{\repeater}[2]
  {\draw[very thick] (#1-.25,#2-.25) rectangle (#1+.25,#2+.25);
\draw[very thick] (#1+.1,#2+.25) -- (#1+.8,#2+.5);
\draw[very thick]  (#1+.8,#2+.5) -- ++(0,0.2);
\draw[very thick]  (#1+.8,#2+.7) -- ++(0.25,0.25);
\draw[very thick]  (#1+.8,#2+.7) -- ++(-0.25,0.25);

\draw[very thick] (#1-.1,#2+.25) -- (#1-.8,#2+.5);
\draw[very thick]  (#1-.8,#2+.5) -- ++(-0,0.2);
\draw[very thick]  (#1-.8,#2+.7) -- ++(-0.25,0.25);
\draw[very thick]  (#1-.8,#2+.7) -- ++(0.25,0.25);
}

\def\bA{{\boldsymbol{A}}}
\def\bB{{\boldsymbol{B}}}

\def\bF{{\boldsymbol{F}}}

\def\bG{{\boldsymbol{G}}}
\def\bH{{\boldsymbol{H}}}
\def\bI{{\boldsymbol{I}}}

\def\bR{{\boldsymbol{R}}}

\def\bT{{\boldsymbol{T}}}

\def\bW{{\boldsymbol{W}}}
\def\bX{{\boldsymbol{X}}}
\def\bY{{\boldsymbol{Y}}}
\def\bZ{{\boldsymbol{Z}}}

\def\bg{{\boldsymbol{g}}}
\def\bh{{\boldsymbol{h}}}

\def\bu{{\boldsymbol{1}}}

\def\bx{{\boldsymbol{x}}}

\def\mS{{{\mathcal{S}}}}

\newcommand{\pp}[1]{{\left( #1 \right)}}
\newcommand{\ppb}[1]{{\left[ #1 \right]}}
\newcommand{\tr}[1]{{\mbox{Tr} \left\{ #1 \right\}}}
\newcommand{\norm}[1]{{ \left\Vert #1 \right\Vert }}
\newcommand{\snorm}[1]{{ \left\Vert #1 \right\Vert^2 }}

\renewcommand{\vec}[1]{{\mbox{vec} \left\{ #1 \right\}}}

\usepackage[color=black,opacity=1]{background}

\SetBgContents{ \begin{minipage}{35cm}{\copyright 2024 IEEE. Personal use
      of this material is permitted. Permission from IEEE must be obtained
       for all other uses, in any current or future media, including
        reprinting/republishing this material for advertising or promotional purposes,
         creating new collective works, for resale or redistribution to servers or lists, 
         or reuse of any copyrighted component of this work in other works. This paper is published  in the IEEE Wireless
         Communications Letters, 2024.}\end{minipage}}
\SetBgScale{.5}
\SetBgAngle{0}
\SetBgPosition{current page.south west}
\SetBgHshift{20cm}
\SetBgVshift{2cm}

\begin{document}
 
\title{Reciprocity Calibration of Dual-Antenna Repeaters}

\author{Erik G. Larsson$^*$, Joao Vieira$^\dagger$, and P\aa l Frenger$^\ddag$
\thanks{This work was supported by the European Union’s Horizon 2020 research and innovation programme under grant agreement No 101013425 (REINDEER), the KAW Foundation, and ELLIIT.
 $^*$Link\"oping University, Dept. of Electrical Engineering (ISY),  581 83 Link\"oping, Sweden, {erik.g.larsson@liu.se}. 
 $^\dagger$Ericsson Research, Mobilv\"agen 12, 223 62 Lund, Sweden,   {joao.vieira@ericsson.com}. 
 $^\ddag$Ericsson Research, Datalinjen 4, 581 12 Link\"oping, Sweden, {pal.frenger@ericsson.com}. 
 }}
 
\maketitle
 
\begin{abstract}
We present a reciprocity calibration method for dual-antenna repeaters
in wireless networks.  The method uses bi-directional measurements
between two network nodes, A and B, where for each bi-directional
measurement, the repeaters are configured in different states.  The
nodes A and B could be two access points in a distributed MIMO
system, or they could be a base station and a mobile user terminal,
for example.  From the calibration measurements, the differences between 
the repeaters' forward and reverse gains are estimated. The repeaters are
then (re-)configured to compensate for these differences such that the repeaters appear,
transparently to the network, as reciprocal components of the propagation environment,
enabling reciprocity-based beamforming in the network.
\end{abstract}


\IEEEpeerreviewmaketitle

\section{Introduction}

\subsection{Repeaters}

In wireless systems, most resources are typically used to serve
disadvantaged users that have low path gains to the base station
(access point).  Such users may be in a shadowed area of the cell, or
 located indoors while the base station is located
outdoors.  In addition, such users typically see channels with low, or
even unit, rank -- prohibiting the transmission of more than a single
data stream.

One technique for improving service to disadvantaged
users is to use \emph{repeaters}
that amplify and instantaneously re-transmit the signal
\cite{patwary2005capacity,sharma2015repeater,garcia2007enhanced,tsai2010capacity,ayoubi2023network,leone2022towards,ma2015channel}.
Repeaters, also known as full-duplex relays, have a small form-factor, and are relatively inexpensive to
build and deploy. They pose, unlike access points in distributed
MIMO, no requirements on phase coherency between geographically
separated units.

\emph{Single-antenna repeaters} use the same antenna for reception and
transmission, and require a circulator to isolate the antenna, the
transmitter port, and the receiver port -- something considered
challenging to implement.  \emph{Dual-antenna repeaters} have two
antenna ports, one for transmission and one for reception.  They have
been standardized in 3GPP, in the form of network-controlled repeaters
\cite{3gpp_repeater}, and typically have a donor (pickup) antenna
outdoors and an antenna indoors to provide coverage.

Our focus is on dual-antenna repeaters.  These repeaters have one
antenna that transmits on uplink and receives on downlink, and one
that does the opposite, regardless of the duplexing mode.
The first antenna is linked, via a forward path with amplification, to the second antenna;
the second antenna is linked, via a reverse path  with amplification, to the first antenna.
In TDD operation, the roles of the repeater's antennas alternate over
time according to the network's TDD pattern.  The repeater
implementations we envision introduce no appreciable delay.  All what
is required is an amplifier circuit in either direction, a tunable
amplitude- and phase-compensation RF circuit, and a control channel
connection (in-band or in a separate band). Also, unless its antennas
are separated sufficiently far apart, the repeater would need either
internal echo cancelation circuitry, or a switch synchronized with the
TDD pattern.

\subsection{Reciprocity and Reciprocity Calibration}\label{sec:rep}

When operating a MIMO system in TDD, the reciprocity of the
uplink-downlink channels facilitates the use of uplink pilots to
acquire downlink channel state information.

The reciprocity principle for electromagnetic wave propagation states
that for two antennas $A_1$ and $A_2$ in a linear-time invariant
medium, a voltage at $A_2$ due to a current source at $A_1$ equals the
voltage at $A_1$ due to the same current source at $A_2$
\cite[p.~119]{harrington2001time}.  When each of $A_1$ and $A_2$ is
connected to a circuit, one then defines the transmitted and received
 {signals}, in both directions, as voltages at specified points in
these circuits.  The relation between these signals constitutes the
[communication-theoretic] channel and it is also reciprocal (equal in both directions);
in some cases (strong coupling, non-identical antennas), 
a linear transformation that depends on the circuit impedances is first required to guarantee that
\cite{laas2020reciprocity}.

In practice, the transmitted and received signals will be affected by
analog components whose phase-lag is slightly temperature- and
age-dependent.  Also, if different antennas are driven
by different oscillators that are not phase-locked (common in
distributed MIMO), their phases will drift apart over time.
These two phenomena  cause a non-reciprocity effect 
that is  well modeled, for each antenna (indexed by $n$, say), by a
time-varying complex-valued coefficient per receive and transmit
branch, herein termed reciprocity coefficient and
denoted by $r_n$ and $t_n$.
These coefficients must be tracked, and compensated for, to maintain
uplink-downlink channel reciprocity; herein, this is called reciprocity calibration.  The modeling with two
coefficients per antenna branch is an approximation that assumes
negligible cross-branch leakage, but has strong experimental support; see, e.g., 
\cite{vieira2017reciprocity,jiang2018framework,shepard2012argos}.

Said non-reciprocity is present in base stations transceivers. It may also be
present in dual-antenna repeaters -- causing their forward and reverse
gains to differ (note that their signal gain depends on the
momentary link direction).  For base stations, many schemes for
estimation of $\{r_n,t_n\}$ exist; they typically rely on
bi-directional over-the-air inter-antenna measurements, see, e.g.,
\cite{vieira2017reciprocity,Kaltenberger,shepard2012argos,lee2017calibration,jiang2018framework}
(for co-located MIMO) \cite{vieralarsson_pimrc,larsson2023phase} (for
distributed MIMO).  None of those methods can be directly used for
reciprocity calibration of repeaters, however.

\subsection{Paper Contribution: Reciprocity-Calibration of Repeaters}

We devise a method of calibrating nominally non-reciprocal
dual-antenna repeaters,  estimating the relation between
their  forward and reverse path gains (to be denoted $\alpha$
and $\beta$).  Knowing this relation, the repeaters can be
re-configured by sending them instructions  over a control
channel, such that their forward and reverse gains  become equal.
This   causes the repeaters to appear as 
reciprocal parts of the propagation environment, entirely transparent
to base stations and the users, making them useful
network components in multiuser TDD systems that employ  reciprocity-based
beamforming.  
 
To the authors' knowledge, there is no previous work on reciprocity
calibration of repeaters.  But it is worth pointing out that previous
literature has identified the non-reciprocity of repeaters to be an
important limitation of their potential to improve system performance
\cite{ma2015channel}.
Somewhat related is  \cite{nie2020relaying}; the authors consider the impact of reciprocity errors
 of   relays (not repeaters),
for which the system model is different.

\begin{figure}[t!]
\begin{center}
\begin{tikzpicture}[xscale=.7,yscale=.9,font=\scriptsize]
 \arrayantenna{0}{0};  
\draw node[anchor=north,align=center] at (.5,-0.5)  {array A \\ $M_A$, $\bR_A$, $\bT_A$ };
\arrayantenna{10}{0};
\draw node[anchor=north,align=center] at (10.5,-0.5)  {array B \\  $M_B$, $\bR_B$, $\bT_B$ };

\draw[semithick,->] plot [smooth] coordinates { (0.5,0.8) (5.5,1.2) (10.5,0.8) };
\draw node[anchor=south] at (5.5,1.2)  {$\bG$};

\draw[semithick,<-] plot [smooth] coordinates { (0.5,0.5) (5.5,0.9) (10.5,0.5) };
\draw node[anchor=north] at (5.5,0.9)  {$\bG^T$};

\repeater{5.5}{-2.1};
\draw node[align=center] at    (5.5,-1.0)  {repeater  \\(R)};

\draw[semithick,dashed,->] plot [smooth] coordinates { (4.4,-1.4) (4.4,-2.3) (5.5,-2.5) (6.6,-2.3) (6.6,-1.4) }; 
\draw[semithick,dashed,<-] plot [smooth] coordinates { (4.2,-1.3) (3.8,-2.4) (5.5,-2.6) (7.2,-2.4) (6.8,-1.3) }; 

\draw node[anchor=east] at (3.8,-2.4)  {$\beta$};
\draw node[anchor=east] at (4.4,-2.3)  {$\alpha$};

\draw[semithick,->] (1.2,-.5) -- (4.4,-1.0) node[pos=0.5,anchor=north] {$\bh^T$}; 
\draw[semithick,<-] (9.8,-.5) -- (6.6,-1.0) node[pos=0.5,anchor=north] {$\bg$}; 

\draw[semithick,<-] (1.2,-.2) -- (4.4,-0.7) node[pos=0.5,anchor=south] {$\bh$}; 
\draw[semithick,->] (9.8,-.2) -- (6.6,-0.7) node[pos=0.5,anchor=south] {$\bg^T$}; 

\end{tikzpicture}
\end{center}
\caption{Two antenna arrays, A and B, and a repeater (R).  By
  definition, $\bG$ is the propagation channel from A to B with R
  turned off.  Radio channels are represented by solid lines ($-$) and
  repeater gains by dashed lines ($--$).\label{fig:1}}
\end{figure}
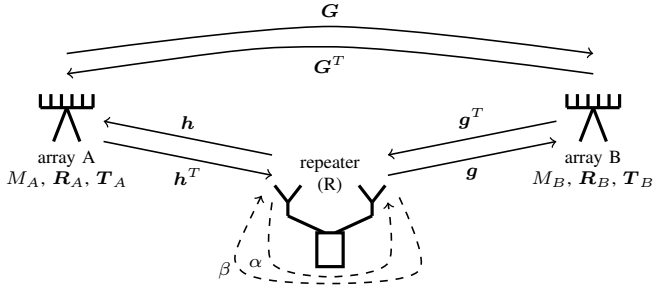

\section{System Model}

We consider a system with two antenna arrays, A and B, and a
dual-antenna repeater, R.  See Figure~\ref{fig:1}. The arrays A and B
have $M_A$ and $M_B$ antennas, respectively. In practice, A and B
could be be a base station and a mobile, or two access points, for
example.

We denote the $M_B\times M_A$ propagation channel from A to B, when R
is turned off, by $\bG$. Reciprocity of propagation holds so the
channel from B to A is $\bG^T$.  No other assumptions are made on the
$\bG$; it could include both line-of-sight and multipath components.
Note that $\bG$ includes the ``passive'' effects of R: even when it is
turned off, R may scatter impinging waves and behave as a
(substantially) reciprocal entity.  We model the non-reciprocity of
the transmit and receive branches at A and B via four diagonal
matrices: $\bR_A$, $\bT_A$, $\bR_B$, and $\bT_B$, comprising the
reciprocity coefficients $\{r_{An}, t_{An}, r_{Bn}, t_{Bn}\}$
introduced in Section~\ref{sec:rep}.

We denote the $1 \times M_A$ propagation channel from A to R by
$\bh^T$, and the $M_A\times 1$ reverse channel from R to A by
$\bh$.\footnote{By convention, throughout, all vectors are column
vectors.}  Similarly, we denote the channel from B to R by $\bg^T$ and
the reverse channel by $\bg$.  When R is in its nominal configuration,
we denote the (complex-valued) gains of its forward and reverse paths
by $\alpha$ and $\beta$.  Generally, $\alpha\neq\beta$; the repeater may nominally not be reciprocal.
 
Taken together, including the reciprocity coefficients in the model,
the (noise-free) channel from A to B is
${ \bR_B  (\bG+\alpha \bg\bh^T) \bT_A} $
and the channel from B to A is
${  \bR_A (\bG^T+\beta \bh\bg^T) \bT_B}. $

For generality, none of the arrays, A or B, is assumed to be a priori
reciprocity-calibrated.  Hence, $\alpha$, $\beta$, $\bG$, $\bg$ and
$\bh$, $\bR_A$, $\bT_A$, $\bR_B$ and $\bT_B$ are a priori unknown.

\section{Calibration Scheme}\label{sec:algs}

Our scheme takes two bi-directional measurements between A and B: (1)
one uni-directional measurement from A to B along with one from B to A with \emph{R in a nominal
configuration}; and (2) one uni-directional measurement from A to B along with one from B to A with
\emph{R configured to rotate the phase of its forward and reverse
gains by $\pi$}.  These measurements are taken within a single channel coherence time interval, by transmitting
pre-determined pilots in a standard manner.  
This gives the four channel estimates,
\begin{align}
\bX_{AB}^0 & = \bR_B (\bG+\alpha \bg\bh^T) \bT_A + \bW_B^0, \quad (M_B\times M_A) \label{eq:mea1} \\
\bX_{BA}^0 & = \bR_A (\bG^T+\beta \bh\bg^T) \bT_B + \bW_A^0, \quad (M_A\times M_B) \label{eq:mea2} \\
\bX_{AB}^1 & = \bR_B (\bG-\alpha \bg\bh^T) \bT_A +  \bW_B^1, \quad (M_B\times M_A)  \label{eq:mea3} \\
\bX_{BA}^1 & = \bR_A (\bG^T-\beta \bh\bg^T) \bT_B +  \bW_A^1, \quad (M_A\times M_B) \label{eq:mea4}
\end{align}
where $\bW_\times^\times$ represents measurement noise.
 
The objective is to estimate $\alpha$ and $\beta$, so all other unknowns are nuisance parameters --
parameters present in the model but of no interest.  The problem of
estimating $\alpha$ and $\beta$ turns out to be non-identifiable, but
the ratio $\beta/\alpha$ can be estimated and this is sufficient   to calibrate the repeater in order to make it   reciprocal.  
Specifically, once $\beta/\alpha$ has been estimated, the repeater can be instructed to
adjust its forward or reverse path gains such that after the
adjustment, these gains are equal.

\subsection{An Ingenuous Approach to Estimation of $\beta/\alpha$}\label{sec:simplealg}

Absent noise, we have from (\ref{eq:mea1})--(\ref{eq:mea4}) that,
\begin{small}
\begin{align}
 2\bX_{AB}^0  \oslash \pp{\bX_{AB}^0+\bX_{AB}^1}  - \bu\bu^T & = -\alpha (\bg \bh^T )\oslash \bG, \label{eq:j1} \\
 \ppb{2\bX_{BA}^0  \oslash \pp{\bX_{BA}^0+\bX_{BA}^1}  - \bu\bu^T }^T  & = - \beta (\bg\bh^T )\oslash \bG  , \label{eq:j2}
\end{align}
 \end{small}
\hspace*{-5mm} where $\oslash$ denotes element-wise (Hadamard) division and ${\bu=[1,\ldots,1]^T}$.  
Dividing (\ref{eq:j2}) by (\ref{eq:j1})  we see that
\begin{small}
\begin{align}\label{eq:simplerat}
&     \ppb{  {2}\bX_{BA}^0  \oslash \pp{\bX_{BA}^0+\bX_{BA}^1}  - \bu\bu^T }^T  \nonumber
    \\ \oslash & \ppb{  {2}\bX_{AB}^0  \oslash \pp{\bX_{AB}^0+\bX_{AB}^1}- \bu\bu^T  }= \frac{\beta}{\alpha} \bu\bu^T.
\end{align}
 \end{small}
\hspace*{-2mm}This observation suggests that one could estimate $\beta/\alpha$ by
averaging the elements of the left hand side of (\ref{eq:simplerat}).
The resulting estimate gives the correct result in the noise-free
case, but it is statistically unsound because of the division by
$\bX_{BA}^0+\bX_{BA}^1$, which can be small.  For example, suppose
$\bG$ is Rayleigh fading with i.i.d.~$CN(0,1)$ elements, and that
there is no noise. Then the elements of $\bX_{BA}^0 \oslash
\pp{\bX_{BA}^0+\bX_{BA}^1} $ have undefined moments.\footnote{To see
this, let $x$ and $y$ be independent $CN(0,1)$.  Then $|x|$ and $|y|$
are independent Rayleigh, and $|x/y|\ge 1/|y|$ with positive
probability. A direct calculation shows that
$\mbox{var}[1/y]=\infty$, so $\mbox{var}[x/y]=\infty$.}  As as consequence, the
performance of this estimator will be erratic, and cannot even be
determined reliably via Monte-Carlo simulation. So while this approach
could appear tempting, it should not be used.  The only purpose of
discussing it here is to make this point.

\subsection{Non-Linear Least Squares (NLS) Criterion}\label{sec:nnls}

To tackle the estimation problem systematically, we start by
re-parameterizing the problem, defining,
\begin{align}\label{eq:reparam}
\bH & = \bR_B \bG \bT_A , & \bA & = \bT_A^{-1} \bR_A , & \bB & = \bT_B \bR_B^{-1} ,\\
\bZ & = \alpha \bR_B \bg \bh^T \bT_A , & \gamma & = \frac{\beta}{\alpha}  .
\end{align}
This re-parameterization is parsimonious, up to an irrelevant scaling
ambiguity -- as will be clear shortly.  The channel estimates,
expressed in the new variables, are
\begin{align}
\bX_{AB}^0 & = \bH +\bZ+ \bW_B^0, \label{eq:mea1e}  \\
\bX_{BA}^0 & = \bA(\bH+\gamma\bZ)^T\bB   + \bW_A^0,   \label{eq:mea2e} \\
\bX_{AB}^1 & = \bH - \bZ  +  \bW_B^1,   \label{eq:mea3e} \\
\bX_{BA}^1 & = \bA(\bH - \gamma \bZ)^T \bB  +  \bW_A^1.   \label{eq:mea4e}
\end{align}

A least-squares estimator, which also yields the maximum-likelihood
solution if all noise components are i.i.d.\ zero-mean Gaussian,
entails minimizing the criterion,
\begin{align}\label{eq:lscrit}
f & = \snorm{ \bX_A^0 - {\bH}  -\bZ } + \snorm{ \bX_B^0 - \bA (\bH+\gamma\bZ)^T\bB } \nonumber \\
+ & \snorm{ \bX_A^1 - ({\bH} - \bZ) } + \snorm{ \bX_B^1 - \bA({\bH}-\gamma \bZ)^T\bB} 
\end{align}
with respect to $\bH$, $\bA$, $\bB$, $\bZ$ and $\gamma$, subject to
the constraints that $\bA$ and $\bB$ are diagonal, and that $\bZ$ has
unit rank.  Hereafter, $\Vert\cdot\Vert$ denotes the Frobenius norm.
The above-mentioned ambiguity is now obvious: multiplication of $\bA$
by an arbitrary constant and division of $\bB$ by the same constant
yields no change in the objective (\ref{eq:lscrit}). But apart from
this ambiguity, all parameters are identifiable.

\subsection{Basic NLS Fitting Algorithm}\label{sec:baseNLS}

A straightforward calculation shows that (\ref{eq:lscrit}) 
can be equivalently written as  
\begin{align}\label{eq:lscrite}
f & = \snorm{ \bR_1 - \bH } + \snorm{ \bR_2 - \bZ} + \snorm{\bR_3 - \bA\bH^T\bB} \nonumber \\
& \qquad + \snorm{ \bR_4 - \gamma \bA\bZ^T\bB},
\end{align}
where we defined the following pre-processed measurements:
\begin{align*}
\bR_1 & = \frac{1}{2}\pp{\bX_{AB}^0 + \bX_{AB}^1} , & \bR_2 & = \frac{1}{2}\pp{\bX_{AB}^0 - \bX_{AB}^1 }, \\
\bR_3 & = \frac{1}{2}\pp{\bX_{BA}^0 + \bX_{BA}^1} , & \bR_4 & = \frac{1}{2}\pp{\bX_{BA}^0 - \bX_{BA}^1 }.
\end{align*}

To approximately find the minimizers of (\ref{eq:lscrite}) we perform
the following sequential algorithmic steps:
\begin{enumerate}
\item Minimize the first term of (\ref{eq:lscrite}) with respect  to ${\bH}$:
\begin{align}
\hat {\bH} = \bR_1.
\end{align}

\item Minimize the second term of (\ref{eq:lscrite}) with
  respect to $\bZ$. This gives
\begin{align}
\hat \bZ =   \mS\{\bR_2 \}, 
\end{align}
where $\mS\{\cdot\}$ is the best rank-one approximation of a matrix in
the least-squares sense (the dominant term in the singular-value
decomposition).

\item Insert $\hat {\bH}$ for ${\bH}$ into (\ref{eq:lscrite}), and
  minimize the third term with respect to $\bA$ and $\bB$. This can be
  done by alternating projection with respect to $\bA$ and
  $\bB$. Specifically, initialize $\hat \bA=\hat
  \bB=\bI$.\footnote{Other initializations are possible.  For example,
  if there is no noise then ${(\bX_{BA}^0 + \bX_{BA}^1)^T \oslash
    (\bX_{AB}^0+\bX_{AB}^1)}$ has rank one, and its left and right
  singular vectors are proportional to the diagonals of $\bB$ and
  $\bA^*$. These could be taken as initial point. However, as noted in
  Section~\ref{sec:simplealg}, element-wise division should be
  avoided.}  Then iterate the following two steps, a pre-determined
  number of steps or until the decrease in the objective ${\norm{\bR_3
      - \bA\bH^T\bB}}$ falls below a threshold:
\begin{align*} 
\hat \bA_{ii} & = \frac{ (\hat \bB\hat {\bH})_{i}^H (\bR_3^{T})_{i}} { \snorm{(\hat \bB\hat {\bH})_{i}  }}, & 
\hat \bB_{ii} & = \frac{ (\hat \bA\hat {\bH}^T)_{i}^H (\bR_3)_{i}} { \snorm{ (\hat \bA\hat {\bH}^T)_{i} }},
\end{align*}
for all $i$, where $(\cdot)_{i}$ denotes the $i$th column of a matrix.
In this iteration, the objective is non-increasing, and non-negative,
so it converges. Because of the scalar ambiguity between $\bA$ and
$\bB$, to ensure numerical stability (avoid that $\hat\bA$ shrinks and
$\hat\bB$ grows, or vice versa), it is sound practice to normalize
$\hat \bA$ and $\hat \bB$ in every iteration, for example, such that
${\snorm{\bA}=1}$.

\item Insert $\hat {\bH}$, $\hat \bA$,
  $\hat \bB$ and $\hat \bZ$ into (\ref{eq:lscrite}), and minimize the
  last term with respect to $\gamma$:
\begin{align}\label{eq:gammahatNLS}
 \hat \gamma = \frac{\tr{(\hat \bA \hat \bZ^T \hat \bB)^H \bR_4 }}
      {\snorm{\hat \bA \hat \bZ^T \hat \bB }}.
\end{align}

\end{enumerate}

\subsection{Improving the Estimate by Alternating Optimization}\label{sec:altNLS}

One can improve the above estimate by alternating optimization,
iterating the following steps:
\begin{enumerate}
\item Insert $\hat\bA$, $\hat\bB$, $\hat\bZ$ and $\hat\gamma$, and minimize   (\ref{eq:lscrite})
with respect to ${\bH}$. 
Letting
\begin{align*}
  \bx  = \begin{bmatrix}
\vec{\bR_1} \\
 \vec{\bR_3^T} 
\end{bmatrix}, 
\ \ \bF  = 
\begin{bmatrix}
\bI_{M_A}\otimes \bI_{M_B} \\
 \hat \bA\otimes \hat \bB
\end{bmatrix},
\end{align*}
where $\otimes$ denotes the Kronecker product, and $\vec{\cdot}$ denotes the stacking of the columns of a matrix on top of one another,
we obtain  $$ \vec{\hat {\bH}} = (\bF^H\bF)^{-1}\bF^H\bx.$$

\item Insert $\hat{\bH}$ into (\ref{eq:lscrite}) and minimize with
  respect to $\bA$ and $\bB$, by iterating the following steps:
\begin{align*} 
\hat \bA_{ii} & =  \frac{ (\hat \bB\hat {\bH})_i^H (\bR_3^{T})_i  + \hat\gamma^*(\hat \bB\hat {\bZ} )_i^H (\bR_4^{T})_i } 
{ \snorm{(\hat \bB\hat {\bH})_i  } + |\hat\gamma|^2 \snorm{(\hat \bB\hat {\bZ} )_i}}, 
\end{align*}
\begin{align*} 
\hat \bB_{ii} & =  \frac{ (\hat \bA\hat {\bH}^T)_i^H (\bR_3)_i  +  \hat\gamma^* (\hat \bA\hat {\bZ}^T)_i^H (\bR_4)_i} 
{ \snorm{ (\hat \bA\hat {\bH}^T)_i } + |\hat\gamma|^2 \snorm{(\hat \bA\hat {\bZ}^T)_i}}.
\end{align*}

\item Insert $\hat\bA$, $\hat\bB$, $\hat{\bH}$ and $\hat\gamma$ into  (\ref{eq:lscrite}) and minimize
with respect to   $\bZ$. The exact minimum  is elusive to us. We suggest 
an approximate solution given by the best rank-one fit to a  linear combination of $\bR_2$ and
$ \hat\bB^{-1} \bR_4^{T} \hat\bA^{-1}$:
\begin{align} \label{eq:Zhat2}
\hat \bZ = \mS\left\{\frac{\bR_2   +  \hat\gamma^* \hat\bB^{-1} \bR_4^{T} \hat\bA^{-1} }{1+|\hat\gamma|^2}\right\}.
\end{align}

\item Insert $\hat\bA$, $\hat\bB$, $\hat{\bH}$ and $\hat\bZ$ into  (\ref{eq:lscrite}) 
and minimize with respect to  $\gamma$; the resulting $\hat\gamma$ is given by (\ref{eq:gammahatNLS}).

\item Repeat from step 1), a pre-determined number of times or until
  the objective (\ref{eq:lscrite}) no longer decreases.
\end{enumerate}

Our proposed NLS fit is statistically sound as it gives the
maximum-likelihood solution in white Gaussian noise. Better algorithms
for minimizing the objective (\ref{eq:lscrite}) might, however,
exist. In the basic NLS fitting algorithm in
Section~\ref{sec:baseNLS}, the obtained estimates
$(\hat\bH,\hat\bZ,\hat\bA,\hat\bB,\hat\gamma)$ are approximate since
no attempt is made to find the global minimum of (\ref{eq:lscrite}).
However, the algorithm is computationally very simple, and yields
consistent estimates in the case of vanishing noise (to show this,
note that $\hat\bH\to\bH$, $\hat\bZ\to\bZ$, and so forth).  The
alternating optimization in Section~\ref{sec:altNLS} yields estimates
that are closer to the global optimum of (\ref{eq:lscrite}). However,
since $\hat\bZ$ in (\ref{eq:Zhat2}) is an approximate minimizer, there
is no guarantee that the objective is non-increasing in every
iteration. To ensure that the algorithm does not ``derail'', we
terminate the iteration in case the objective would increase (which
can happen because of said approximation).

\section{Numerical Example}

\begin{figure}[t!]
    \centerline{\scalebox{.7}{\input{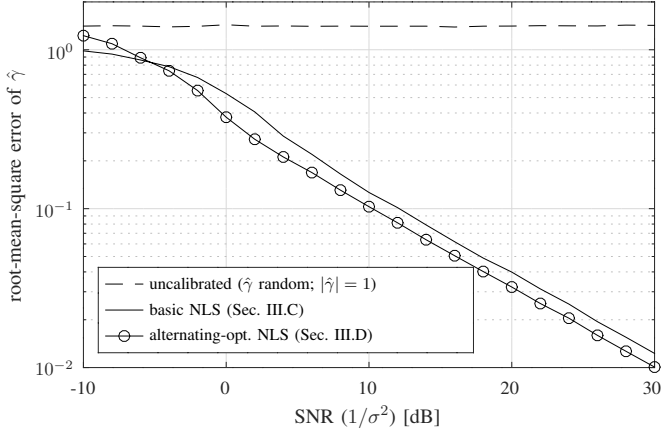}}}
    \caption{Root-mean-square-error of $\hat\gamma$. 
      }
\label{fig:sim}
\end{figure}

We consider an example where A has $M_A=4$ antennas and B has $M_B=3 $
antennas. The vectors $\bh$ and $\bg$ are random columns of
$M_A\times M_A$ and $M_B\times M_B$ DFT matrices,
  modeling line-of-sight propagation between A and R and between R and B. 
The propagation channel between A and B with R off,
$\bG$, has  independent $CN(0,1)$ entries, modeling Rayleigh fading.
The diagonal entries of $\bR_A$, $\bT_A$, $\bR_B$ and $\bT_B$  have unit magnitudes,   
and independent, uniformly random phases  over  $[-\pi,\pi]$.  The noise matrices
$\bW_A^0$, $\bW_B^0$, $\bW_A^1$, and $\bW_B^1$ have independent
$CN(0,\sigma^2)$ elements;   $1/\sigma^2$ is the SNR as measured at any
antenna of B when one antenna at A transmits with unit power and R is
off.  

Figure~\ref{fig:sim} shows the performance of the basic NLS and
alternating-optimization NLS estimators, for repeater gains
$|\alpha|^2=|\beta|^2=10$~dB.  100 iterations were run in the
alternating projection subroutines that minimize with respect to $\bA$ and
$\bB$ (Step 3 in Section~\ref{sec:baseNLS} and Step 2 in
Section~\ref{sec:altNLS}). At high SNR, the root-mean-square error
decreases by approximately a factor 10 when increasing the SNR 20 dB.
This is expected, as for non-linear models with additive white
Gaussian noise, the curvature of the log-likelihood is proportional to
the SNR.  The alternating optimization in Section~\ref{sec:altNLS} was
run for at most 25 iterations; it improves performance almost 2~dB, at the
cost of correspondingly higher complexity.

\section{Variations and Extensions}

\subsection{Pre-Calibrated Arrays}

If A and B have been previously reciprocity-calibrated, for example
using the technique in \cite{vieralarsson_pimrc}, then $\bA$ and $\bB$
can be considered known.  The above algorithm substantially simplifies
in this case, as the only unknowns are 
$\gamma$, $\bH$ and $\bZ$;
we leave the details to the reader. However, we stress that
this requires A and B to be \emph{jointly}
reciprocity-calibrated. Individual calibration is insufficient -- for
an explanation of the difference, see \cite{larsson2023phase}.  For
example, if A and B are driven by independent oscillators, then  
re-calibration of A and B for joint reciprocity must be done every
time the oscillators have drifted apart by more than some predefined
angle.

\subsection{Turning On and Off  Instead of Phase-Shifting}
 
Instead of performing one bi-directional measurement with R in a
nominal configuration and one bi-directional measurement instructing R to phase-shift by $\pi$,
one can take one such measurement with R on and one with R off.  We get the two
bi-directional measurements (ignoring noise henceforth for
simplicity):
\begin{align}
\bY_{AB}^0 & = \bH +\bZ ,  &  
\bY_{BA}^0 & = \bA(\bH+\gamma\bZ)^T\bB , \label{eq:rph1} \\
\bY_{AB}^1 & = \bH  , &   
\bY_{BA}^1 & = \bA \bH^T \bB  . \label{eq:rph2} 
\end{align}
The NLS criterion in (\ref{eq:lscrit}) can be easily modified to this
case,   resulting in similar algorithms as in
Section~\ref{sec:algs}.

\subsection{Simultaneous Calibration of Multiple Repeaters}

With multiple repeaters in the network, their operation needs to be
coordinated during calibration activities.  One can of course,
trivially, calibrate one repeater at a time by turning off all but the
one that is up for calibration.   
However, one can do better by instructing the repeaters to rotate the phase
of their gains according to a pre-determined pattern.  We exemplify
this as follows, using phase rotations of $0$ and $\pi$.  

\clearpage

Consider a setup with $N$ repeaters, $R_1, \ldots, R_N$, with
corresponding rank-one channels $\bZ_1, \ldots, \bZ_N$ and gain ratios
$\gamma_1,\ldots,\gamma_N$.  First, as an initial step, configure the
repeaters to set their gains equal to a nominal value. This yields the
(noise-free) bi-directional measurement
\begin{align}
\bY_{AB}^0  & = \bH+\bZ_1+ \cdots +\bZ_N, \label{eq:multi1} \\
\bY_{BA}^0 & = \bA(\bH+\gamma_1 \bZ_1+\cdots +\gamma_N \bZ_N )^T \bB. \label{eq:multi2}
\end{align}
Next, configure all repeaters to rotate the phase of their gains by $\pi$, resulting in the bi-directional measurement
\begin{align}
\bY_{AB}^1  & = \bH-\bZ_1-\cdots-\bZ_N, \label{eq:multi3} \\
\bY_{BA}^1 & = \bA(\bH-\gamma_1 \bZ_1-\cdots -\gamma_N \bZ_N )^T \bB. \label{eq:multi4}
\end{align}
From (\ref{eq:multi1})--(\ref{eq:multi4}), $\bA\bH$ and $\bB\bH\bA$
can be recovered (up to a multiplicative scalar ambiguity between
$\bA$ and $\bB$). For example, $\bH$ can be estimated by averaging
$\bY_{AB}^0$ and $\bY_{AB}^1$, and $\bB\bH\bA$ can be estimated by
averaging $(\bY_{BA}^0 )^T$ and $(\bY_{BA}^1 )^T$. Consequently, any
terms of the form $\bH$ or $\bB\bH\bA$ that appear in subsequent
measurements can be eliminated. From now on, we assume that such
elimination has been performed. Also, we can assume that the diagonal
matrices $\bA$ and $\bB$ are known (up to a multiplicative scalar
ambiguity) since they can be determined from $\bH$ and $\bB\bH\bA$.

After this initial step, instruct the repeaters to rotate their phases
according to a pre-determined pattern.  We give an example, for $N=4$,
using a pattern from a Hadamard matrix. For each pattern,
bi-directional measurements are taken. This yields, after subtracting
the already-obtained estimates of $\bH$ and $\bB\bH\bA$, and after
transposition:
\begin{align}
& \bZ_1+\bZ_2+\bZ_3+\bZ_4 \label{eq:muria} \\
& \bB(\gamma_1 \bZ_1+\gamma_2 \bZ_2+\gamma_3 \bZ_3+\gamma_4 \bZ_4 )\bA \label{eq:murib} \\
& \bZ_1-\bZ_2+\bZ_3-\bZ_4 \label{eq:muric} \\
& \bB(\gamma_1 \bZ_1-\gamma_2 \bZ_2+\gamma_3 \bZ_3-\gamma_4 \bZ_4 )\bA \label{eq:murid} \\
& \bZ_1+\bZ_2-\bZ_3-\bZ_4 \label{eq:murie} \\
& \bB(\gamma_1 \bZ_1+\gamma_2 \bZ_2-\gamma_3 \bZ_3-\gamma_4 \bZ_4 )\bA \label{eq:murif} \\
& \bZ_1-\bZ_2-\bZ_3+\bZ_4 \label{eq:murig} \\
& \bB(\gamma_1 \bZ_1-\gamma_2 \bZ_2-\gamma_3 \bZ_3+\gamma_4 \bZ_4 )\bA .\label{eq:murih}  
\end{align}
Note that (\ref{eq:muria}) and (\ref{eq:murib}) were already obtained
in the initial step when $\bH$ and $\bB\bH\bA$ were estimated; we
repeat them here for completeness.  Once $\bA$ and $\bB$ are estimated, by
forming linear combinations (\ref{eq:muria})--(\ref{eq:murih}), we can
recover $\{\bZ_n\}$ and $\{\gamma_n \bB\bZ_n \bA\}$ by inverting the
Hadamard pattern, and then recover $\{\gamma_n\}$.  For example,
adding (\ref{eq:muria}), (\ref{eq:muric}), (\ref{eq:murie}), and
(\ref{eq:murig}) yields $4\bZ_1$. Adding (\ref{eq:murib}),
(\ref{eq:murid}), (\ref{eq:murif}), and (\ref{eq:murih}) yields
$4\gamma_1 \bB\bZ_1 \bA$.  Since $\bA$ and $\bB$ are known, $\gamma_1$
can then be found from these two sums.  Similarly, by forming
other linear combinations of (\ref{eq:muria})--(\ref{eq:murih}),
$\gamma_2,\cdots,\gamma_4 $ can be determined.

Other constructions, for example based on DFT matrices, are 
also possible.  Orthonormal patterns are preferable, as they yield a
well-conditioned inversion in the last step.

\section{Concluding Remarks}

The potential of using repeaters to enhance performance of
reciprocity-based multiuser MIMO has been discounted in the past
because their presence, nominally, breaks the uplink-downlink
reciprocity.  With our proposed reciprocity calibration technique,
dual-antenna repeaters can be made into transparent, reciprocal components
of the radio wave propagation environment  -- enabling improved multiplexing
capability and improved coverage to disadvantaged users.

Our scheme requires a reasonable SNR on the direct link (with the
repeater off) to achieve good-quality estimates.  In a deployment with
many repeaters, as long as one can calibrate a first repeater, one
could use it to calibrate a second repeater, even if there is no good
direct-link to calibrate this second repeater, and so on.

\end{document}